\documentclass[aps,twocolumn]{revtex4}

\usepackage{graphicx}                       
\usepackage{amsmath,amsfonts}
\usepackage{bm}                             
\graphicspath{{./Figures/}}                 

\newcommand{\op}{\langle\bar{\psi}i\gamma_{5}\tau^{3}\psi\rangle}  
\newcommand{\gmplane}{\mbox{$(g^2, m_0)$ plane }}                  
\newcommand{\bkplane}{\mbox{$(\beta, \kappa)$ plane }}             
\newcommand{\etal}{\textit{\mbox{et al.\ }}}                       
\newcommand{\ie}{\textit{\mbox{i.e.\ }}}                           
\newcommand{\eg}{\textit{\mbox{e.g.\ }}}                           
\newcommand{\effchi}{$\chi^2/\textsc{ndf}$}                        

\begin{document}


\title{A numerical reinvestigation of the Aoki phase \\
       with \bm{$N_f=2$} Wilson fermions at zero temperature}

\author{E.--M.~Ilgenfritz, W.~Kerler, M.~M\"uller--Preussker, A.~Sternbeck}
\affiliation{Humboldt-Universit\"at zu Berlin, Institut f\"ur Physik, 
             D-12489 Berlin, Germany}
\author{and H.~St\"uben}
\affiliation{Konrad-Zuse-Zentrum f\"ur Informationstechnik Berlin, 
             D-14195 Berlin, Germany}

\date{January 19, 2004}

\begin{abstract}
  We report on a numerical reinvestigation of the Aoki phase in
  lattice QCD with two flavors of Wilson fermions where the
  parity-flavor symmetry is spontaneously broken.  For this purpose
  an explicitly symmetry-breaking source term $h\bar{\psi} i \gamma_{5}
  \tau^{3}\psi$ was added to the fermion action. The order
  parameter $\op$ was computed with the hybrid Monte Carlo algorithm 
  at several values of $(\beta,\kappa,h)$ on lattices of sizes $4^4$ to
  $12^4$ and extrapolated to $h=0$. The existence of a parity-flavor 
  breaking phase can be confirmed at $\beta=4.0$ and $4.3,$ while 
  we do not find parity-flavor breaking at $\beta=4.6$ and $5.0$.
\end{abstract}

\keywords{Wilson fermions, phase diagram, parity-flavor symmetry, Aoki phase}

\maketitle

\section{Introduction} \label{sec:intro}

Spontaneous breaking of chiral symmetry is one of the main
non-perturbative phenomena of QCD explaining many features of the
hadronic world, in particular of hadrons containing $u$, $d$ and/or
$s$ quarks.  QCD allows us to interpret the light octet mesons as Goldstone
bosons.  The four-dimensional (Euclidean) lattice discretization of
QCD provides a unique \emph{ab initio} non-perturbative approach. 
However, in this
approach chiral symmetry has to be treated with special care. At
present, on the lattice this symmetry is best realized by satisfying
the Ginsparg-Wilson relation \cite{GinWil} for the lattice Dirac
operator, \eg\!\!, employing the so-called overlap operator \cite{NarNeu,Neu}, 
or using the five-dimensional domain wall fermion ansatz
\cite{KapSham,Sham,FurSham}.  In both cases the Wilson-Dirac operator $W(m_0)$
(with a bare mass parameter $m_0 \in (-2, 0)$) serves as an input for
the fermionic part of the lattice discretized action.

For the Wilson-Dirac operator (which breaks chiral invariance
explicitly)  Aoki \cite{Aoki_first} has argued that in a certain range of
the hopping parameter $\kappa$ (or the bare mass $m_0$) there is a
phase in which parity-flavor symmetry is spontaneously broken,
in the sense that a condensate as defined in Eq.~(\ref{eq:lim_h_V}) 
exists and is non-vanishing. 
In agreement with the literature we call it the {\it Aoki phase}.  
When $\kappa$ approaches the border lines of this phase
all pion masses tend to zero because one is approaching a second order
phase transition.  In the whole Aoki phase the charged pion states are
expected to remain massless (in the case of $N_f=2$ flavors) since they
appear to be the Goldstone bosons related to parity-flavor breaking,
whereas the neutral pion should become massive again.
The general phase structure as proposed by Aoki is shown in Fig.~1.
Some numerical results supporting this picture were presented in 
Refs.~\cite{Aoki_first,Aoki_next,AokGoc1,AokGoc2,AokKanUka,BitLATT97}.

It has been questioned whether the Aoki phase survives the
continuum limit (in the sense of extending to $\beta = \infty$) or,
alternatively, ends somewhere at finite $\beta$, 
perhaps before the scaling regime is reached.  Previous
investigations of this problem did not yield a unique 
\mbox{answer \cite{AokGoc2,AokKanUka,BitLATT97,ShaSin,BitPRD,Aoki_summary}}.
In this paper we present results of a more thorough
numerical analysis of this question.  As has been discussed
recently \cite{GolSham}, the answer is of relevance for the locality
behavior and the restoration of chiral invariance in quenched and full
QCD with Ginsparg-Wilson and domain wall fermions.  Accordingly, the 
region of the Aoki phase has to be avoided in such computations in 
order not to spoil physical reliability.
  
Our investigation was carried out for full lattice QCD with $N_f=2$
flavors of unimproved Wilson fermions using the standard plaquette
gauge action.  It includes a careful extrapolation of $\op$ to
vanishing external field.  It shows that the Aoki phase is unlikely to
extend beyond $\beta = 4.6$ (which confirms early conclusions in
Ref.~\cite{BitPRD}).

The outline of our paper is as follows.  In Sec.~II we discuss the
proposed phase structure in greater detail. Section~III provides 
details of our numerical simulations.  In Sec.~IV we present our
numerical results. Section~V contains the discussion and our 
conclusions.
  
\section{The proposed phase structure}
\label{sec:Aokiphase} 

\begin{figure*}
  \centering
  \includegraphics[height=5cm]{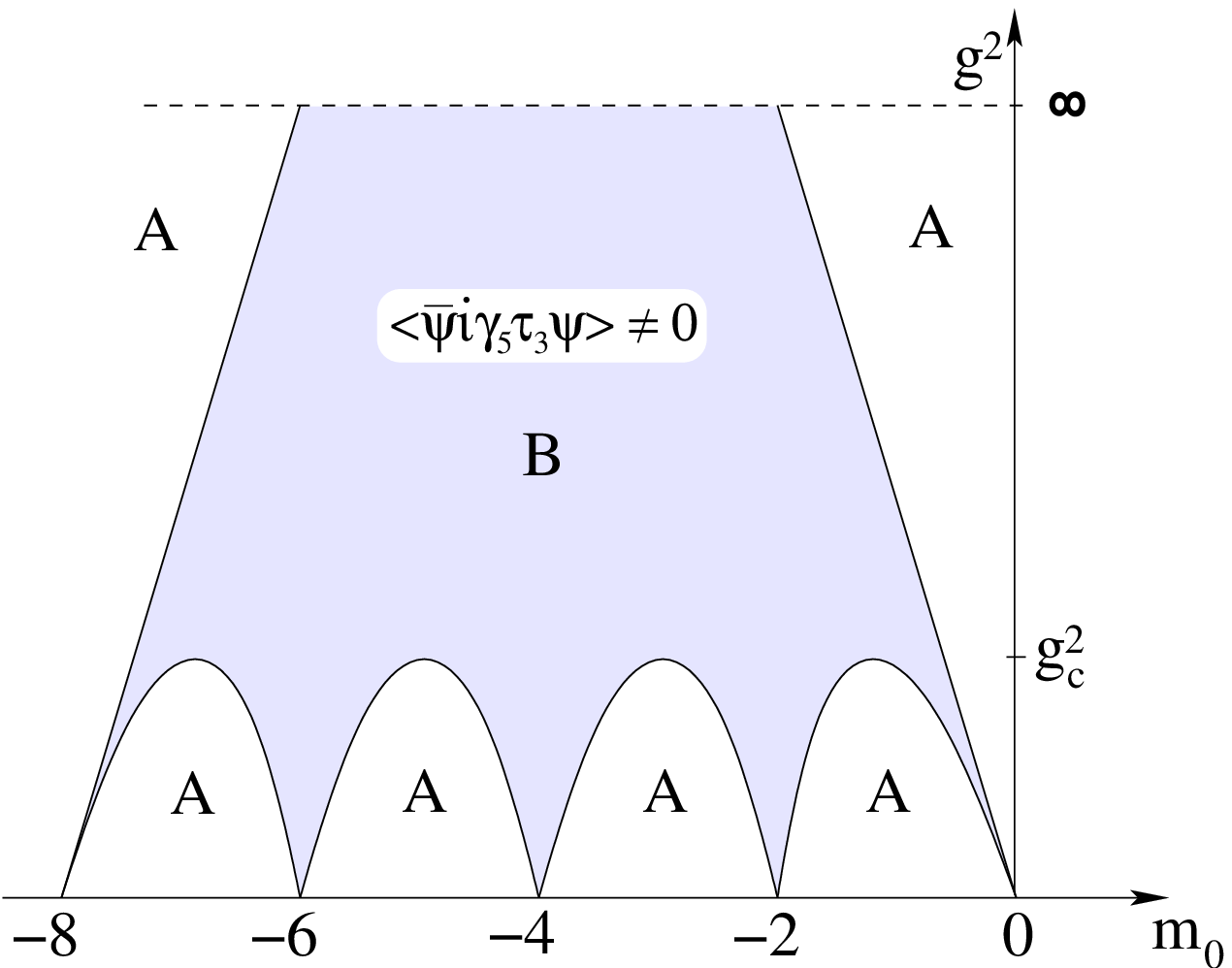}\qquad%
  \includegraphics[height=5cm]{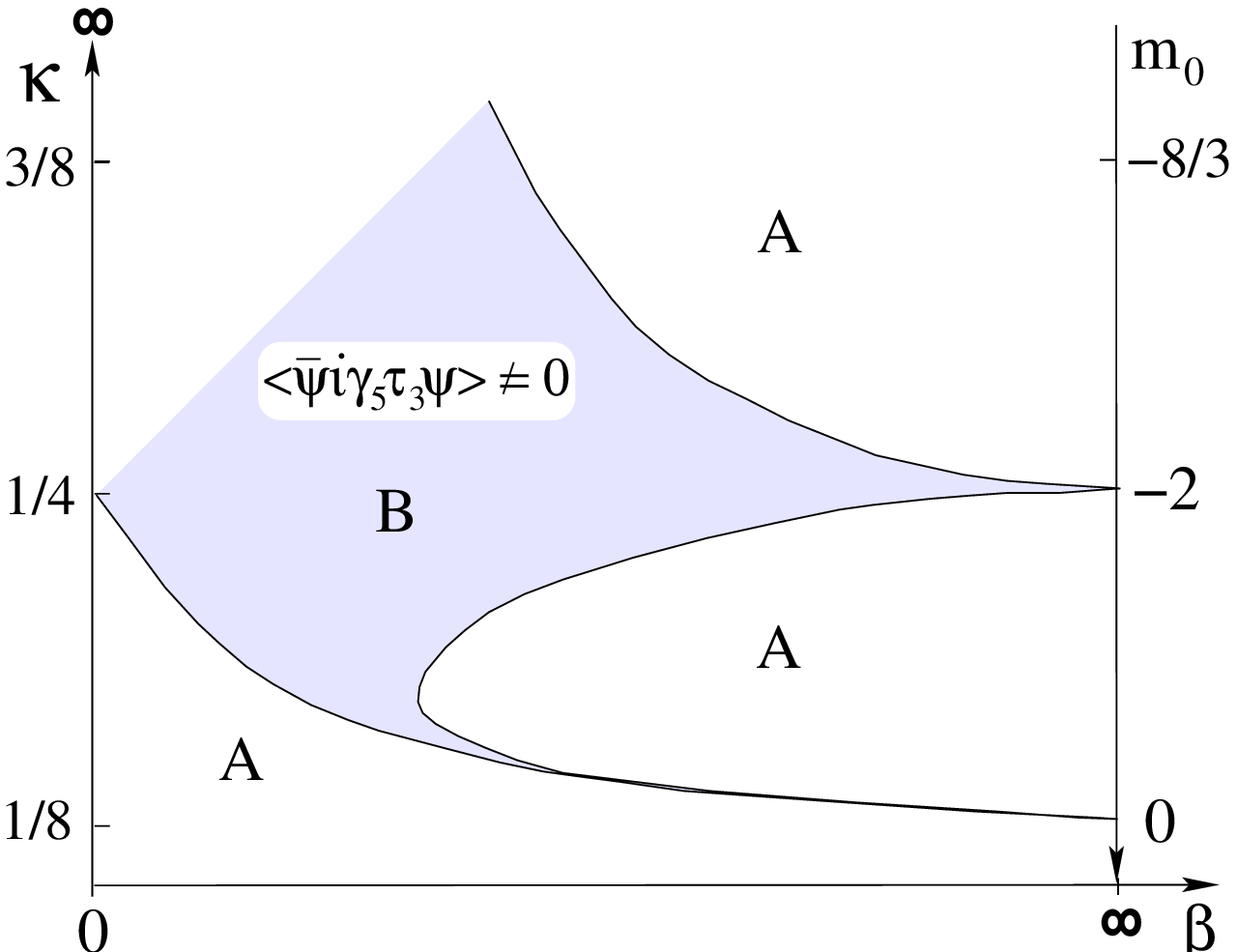}
  \caption{The phase diagram proposed by Aoki \etal in the \gmplane
(left-hand side) and in the \bkplane (right-hand side).  The shaded
region labeled $B$ denotes the phase where flavor and parity are
spontaneously broken.  Both symmetries are conserved in regions
labeled $A$.}
  \label{fig:proposed_phase_diagram}
\end{figure*}

Aoki, in his last status report~\cite{Aoki_summary},
has discussed the lattice results supporting the view
that for lattice QCD with $N_f=2$
Wilson fermions there exists a parity-flavor breaking phase which is
separated from an unbroken phase (or from unbroken phases) by second
order phase transition lines.  
The conjectured phase structure in the \gmplane is shown on the left-hand 
side of Fig.~\ref{fig:proposed_phase_diagram}.  As can be seen from this
figure, two of these critical lines run from strong coupling to the
weak coupling limit, while further critical lines are confined to the
weak coupling region.  At zero coupling, pairs of these transition
lines join at points referring to the different fermion doublers.
Aoki \cite{Aoki_next} has further claimed that along the critical lines 
the pion triplet is massless. The neutral pion becomes massless only on the
critical lines, due to the presence of a second order phase
transition, while the charged pions turn massless on the critical
lines and remain massless inside the Aoki phase signaling that flavor
symmetry is broken.

When simulating the theory it is natural to draw the phase diagram 
in the \bkplane. Using the well known relations $\kappa=1/(2m_{0}+8)$ and
$\beta=6/g^2$, the proposed phase structure is mapped to this plane as
shown on the right-hand side of Fig.~\ref{fig:proposed_phase_diagram}.
Therein the symmetry \mbox{$m_0\leftrightarrow -(m_0+8)$} is hidden in the 
reflection $\kappa \leftrightarrow -\kappa$ which is not 
made explicit for simplicity. The critical line $\kappa_c(\beta)$ 
which runs from $\beta=0$ to infinity is nothing but the chiral limit 
line of lattice QCD. Thus the scenario proposed by Aoki \etal 
might explain why all pions are massless along this line despite 
the fact that Wilson fermions explicitly break chiral symmetry.

In principle, the Aoki phase could be expected to exist for all values
of $\beta$.  In the strong coupling region the existence of such a
phase was verified by performing numerical simulations of QCD with
Wilson fermions as summarized in \cite{Aoki_summary} and reconsidered in
\cite{BitLATT97}.  For this purpose a so-called
{\it twisted mass} term $h\bar{\psi} i \gamma_{5} \tau^{3}\psi$ was
added to the action which explicitly breaks parity-flavor
symmetry.  Without an external field $h$ coupling to $\bar{\psi} i
\gamma_{5} \tau^{3} \psi$, the parameter $\op$ would always
be zero on a finite lattice.  $\op$ has to be measured for
varying lattice size $V$ and non-vanishing $h$ values.  The order
parameter $\op_{h=0}$ is then obtained by taking the double limit
in the following order
\begin{equation}
\label{eq:lim_h_V}
  \op_{h=0}=\lim_{h\rightarrow 0}\lim_{V\rightarrow \infty}
  \langle\bar{\psi} i \gamma_{5} \tau^{3}\psi\rangle \,.
\end{equation}
In the literature one finds numerical results from quenched 
\cite{AokKanUka} and unquenched \cite{AokGoc2,BitLATT97} simulations 
at finite $h$ which support the existence of a parity-flavor breaking phase, 
at least for $\beta\le 4.0$. However, extrapolations in order to carry out
the double limit (\ref{eq:lim_h_V}) had not been performed.

Going to larger values of $\beta$ there are contradictory statements
about the existence of such a broken phase.  Bitar \cite{BitPRD} has
come to the conclusion that there is no Aoki phase for $\beta \ge
5.0$.  However, results from quenched simulations \cite{AokKanUka} suggest 
that the \emph{finger} structure anticipated by Aoki exists.  

Aoki's scenario was also challenged 
in Refs.~\cite{BitHelNar,EdwHelNarSin,EdwHelNar}. In particular,
in Ref.~\cite{BitHelNar} it has been argued that flavor and parity 
are not violated at finite lattice spacing. The authors have rather 
proposed that the Aoki phase has to be interpreted as a phase with 
massless quarks and spontaneous chiral symmetry breaking. 

In Ref.~\cite{ShaSin} the controversy has been concisely elucidated in the 
sense that, at finite lattice spacing $a$, the Wilson lattice theory is
able to exhibit flavor and parity breaking under certain circumstances.
The authors have also demonstrated that the results of 
\cite{EdwHelNarSin,EdwHelNar} 
concerning the spectrum of the Hermitean Wilson-Dirac operator 
$\gamma_5 W(m_0)$ (actually obtained for quenched or partially quenched 
lattice QCD) lend support, if correctly interpreted, to a non-vanishing 
condensate as defined in Eq.~(\ref{eq:lim_h_V}). 

In terms of an effective chiral Lagrangian, it has been pointed out in 
Ref.~\cite{ShaSin} that only two possible scenarios may exist, 
depending on the sign of one coupling coefficient. In the first case, 
the Aoki picture~\cite{Aoki_next} is exactly reproduced, whereas in the second 
case all pion masses remain degenerate and non-vanishing over the whole 
\gmplane such that no Aoki phase exists at all. 
If the first case applies to lattice QCD all the way to the continuum limit 
specific predictions concerning the $a$ dependence of the neutral 
pion mass and of the width of the \emph{Aoki finger} 
pointing towards $(m_0=0,g^2=0)$ have been made. However, 
if the sign turns to the second case the Aoki phase ceases to exist at strong
coupling and those predictions do not apply.

After that, the only remaining question is whether the Aoki phase really 
persists until the continuum limit and, if it does so, how it shrinks to the 
point $(m_0=0,g^2=0)$. 
Only numerical simulations can clarify whether there is a strip 
of parity-flavor breaking phase in lattice QCD with Wilson fermions 
extending to infinite $\beta$.

\begin{figure*}
  \centering
  \vspace{-1cm}
  \includegraphics{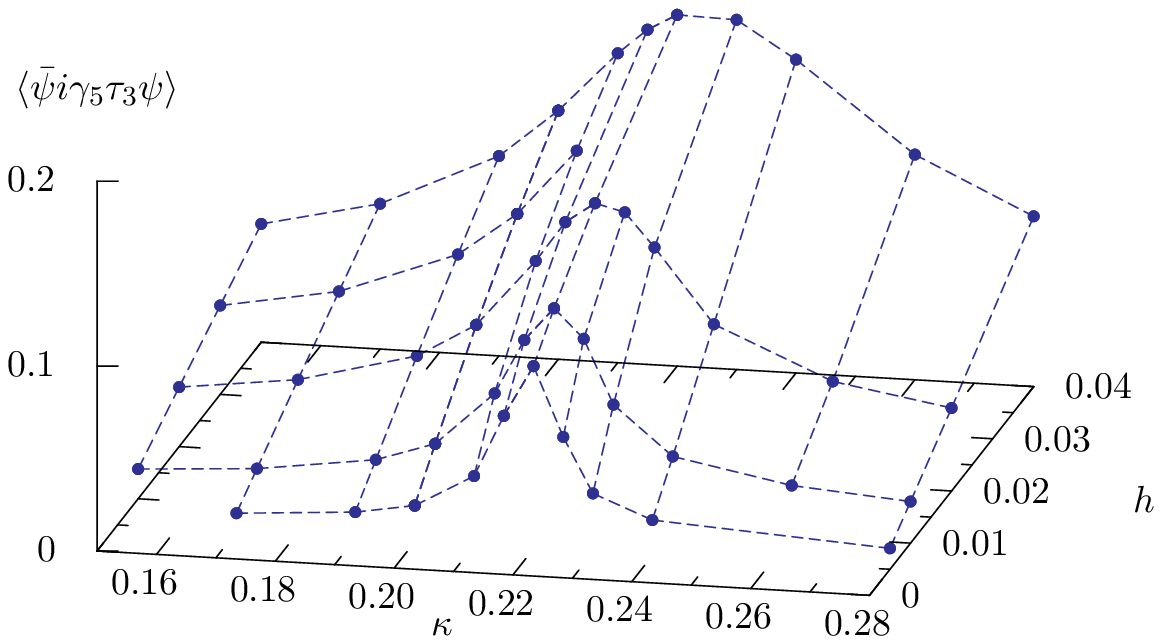}
  \vspace{-1cm}
  \caption{Results for $\op$ as a function of $\kappa$ and $h$ 
    at $\beta=4.0$ on a $6^4$ lattice.}
  \label{fig:orderparameter3D}
\end{figure*}

\section{Simulation details} \label{sec:simulations}

We have simulated lattice QCD with two flavors of (unimproved) Wilson 
fermions with (the $\Phi$ version of) the hybrid Monte Carlo 
algorithm~\cite{Duane,Gottlieb} 
where an even/odd decomposition~\cite{even-odd} 
has been employed. An explicitly
symmetry breaking source term was added to the Wilson fermion matrix
$M_W$, \ie the two-flavor fermion matrix was given by
\begin{equation}
  \label{eq:fermionmatrix}
  M(h)=M_W+ h i\gamma_5\tau^3.
\end{equation}
The simulations were performed on lattices ranging from $4^4$ to
$12^4$ at $\beta$ values 4.0, 4.3, 4.6, and 5.0, with $\kappa$ and $h$ in the
intervals $0.15\le\kappa\le0.28$ and $0.003\le h\le0.04$, respectively.  

In our study we measured $\op$ as a function of $\kappa$ at finite $h$.
The parameter $\op$, which is proportional to the imaginary part of the 
trace of $\gamma_5 M^{-1}(h)$, was averaged over 100--1000
gauge field configurations (see Table \ref{tab:stat}) 
separated by trajectories of length 1.  The
trace was measured with a stochastic estimator \cite{Horsley}.

For illustration, results for $\op$ from a $6^4$ lattice at
$\beta=4.0$ are shown in Fig.~\ref{fig:orderparameter3D}.  
The location of the peak determines the region where 
subsequent simulations on larger
lattices and smaller $h$ were performed.  In
Fig.~\ref{fig:orderparameter3D} the peak is around $\kappa=0.22$. It
becomes sharper as $h$ decreases.  We have increased the lattices
until measurements agreed within errors such that we can treat our
largest lattices as infinitely large.  The extrapolation to vanishing 
$h$ is described in the following section.

\section{Extrapolating to vanishing external field}
\label{sec:extrapolation}

\begin{figure*}
  \centering
  \mbox{\includegraphics[height=8cm]{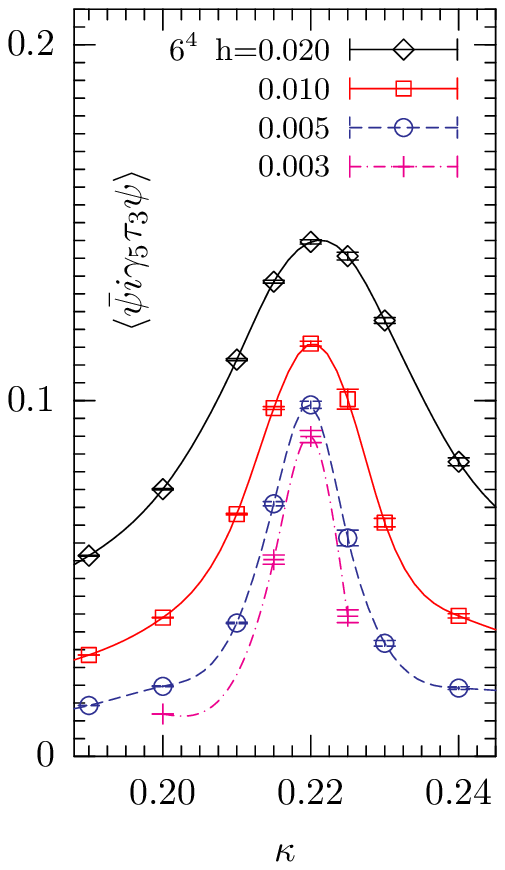}\hspace{-0.9cm}
    \includegraphics[height=8cm]{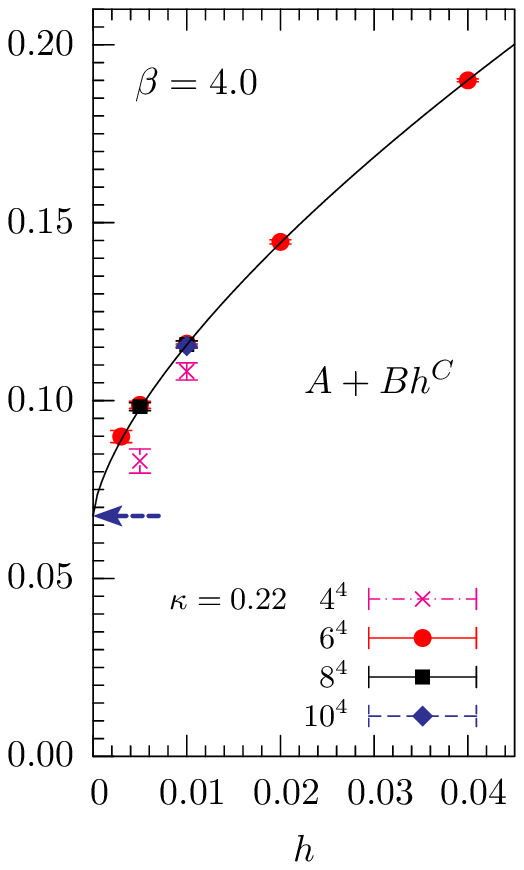}\hspace{-0.7cm}
    \includegraphics[height=8cm]{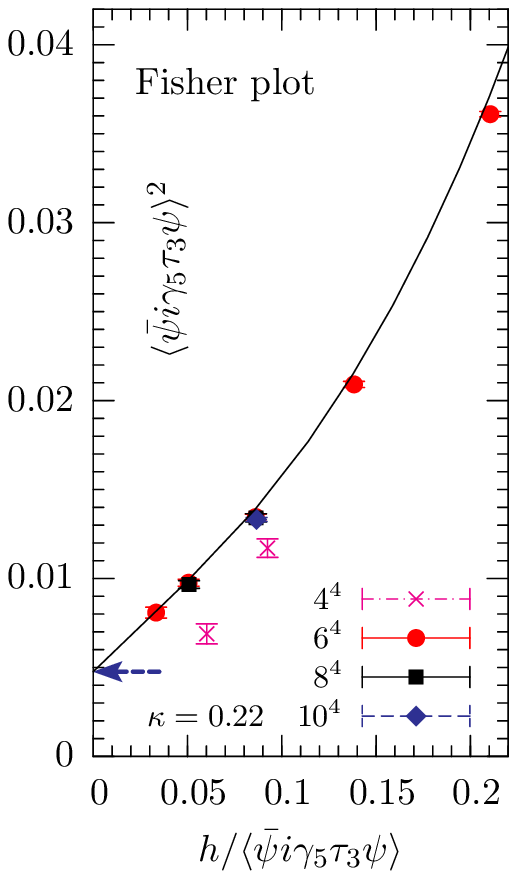}}

  \mbox{\includegraphics[height=8cm]{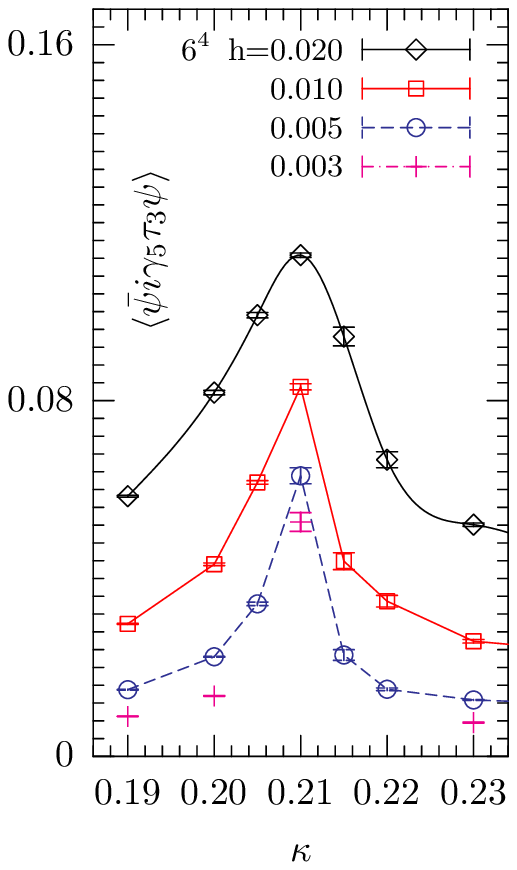}\hspace{-0.9cm}
    \includegraphics[height=8cm]{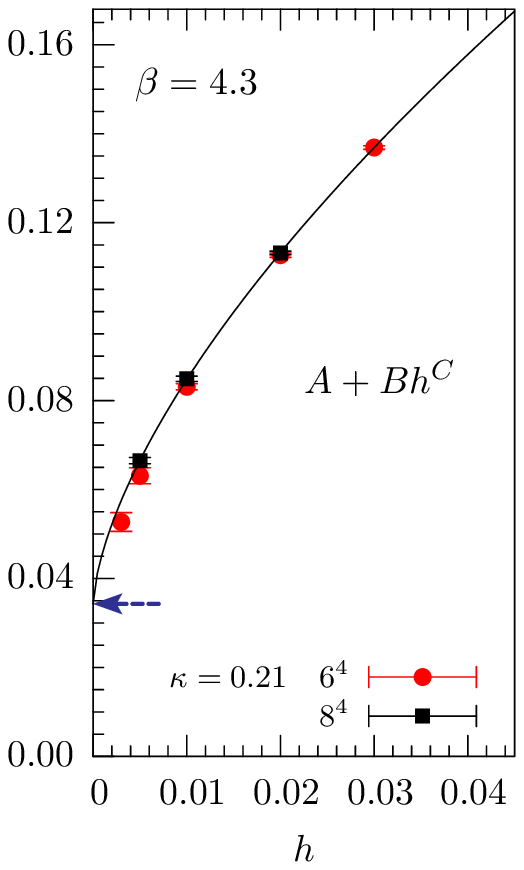}\hspace{-0.7cm}
    \includegraphics[height=8cm]{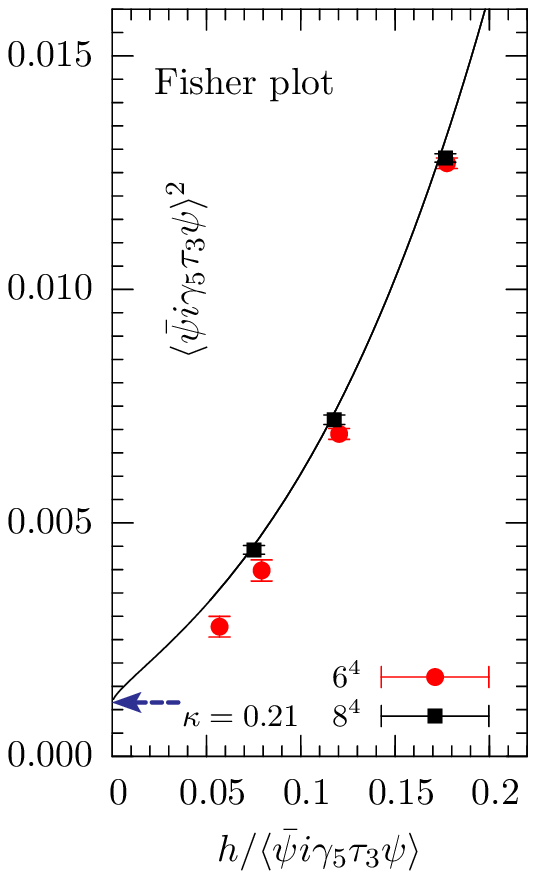}}
  \caption{In the left column data for $\op$ from
  a $6^4$ lattice are shown as a function of $\kappa$ at several
  values of $h$ (the lines are spline interpolations to guide the eye). 
  The extrapolation to $h=0$ in the infinite volume limit 
  is shown in the center column of this figure. 
  The right column shows the Fisher plots with the corresponding fitting
  function. The upper row shows results for $\beta=4.0$, the lower one for 
  $\beta=4.3$.}
\label{fig:extrapolation4043}
\end{figure*}

In Fig.~\ref{fig:extrapolation4043} an analysis of $\op$ data 
is shown for $\beta = 4.0$ and $4.3$.  As can be seen from the
upper and lower left plot the interesting region is around
$\kappa=0.22$ and $\kappa=0.21,$ respectively.  At these
$(\beta,\kappa)$ pairs further simulations were performed in order to
control finite-size effects.  Data from these simulations are shown
in the center plots of Fig.~\ref{fig:extrapolation4043}.  No finite
size effects are visible in the plots except for data from the $4^4$
lattice at $\beta=4.0$.  Hence, the measurements of $\op$ from the
largest lattice at each $h$ can be considered to lie within errors on
the infinite volume envelope.

The question arises how to fit these data properly.  Motivated from
the mean field equation
\begin{equation}
  h= A_0\sigma^3 + A_1\!\cdot\!\left(\kappa-\kappa_c\right)\sigma\qquad
 \textrm{with}\quad\sigma\equiv\op
 \label{eq:meanfield}
\end{equation}
we use the ansatz
\begin{equation}
  \sigma(h) = A + Bh^{C} + \ldots\quad.
  \label{eq:ansatz}
\end{equation}
It is instructive to look at
so-called Fisher plots \cite{Fisher,Hinnerk2} (see the right-hand side
of Fig.~\ref{fig:extrapolation4043}). From Eq.~(\ref{eq:meanfield}) 
one expects data for $\kappa\le\kappa_c$ to lie
on straight lines ending at the origin or at the abscissa, while
within the broken phase they should lie on straight lines ending at
the ordinate.  As can be seen from the Fisher plots obtained the data 
do not lie on straight lines and therefore do not behave mean-field like.

\begin{figure*}
  \centering
  \mbox{\includegraphics[height=8cm]{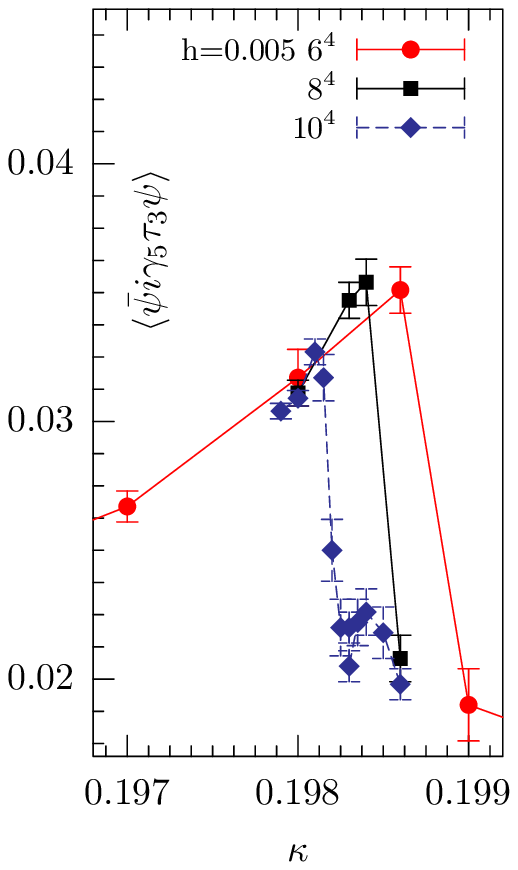}\hspace{-0.5cm}
    \includegraphics[height=8cm]{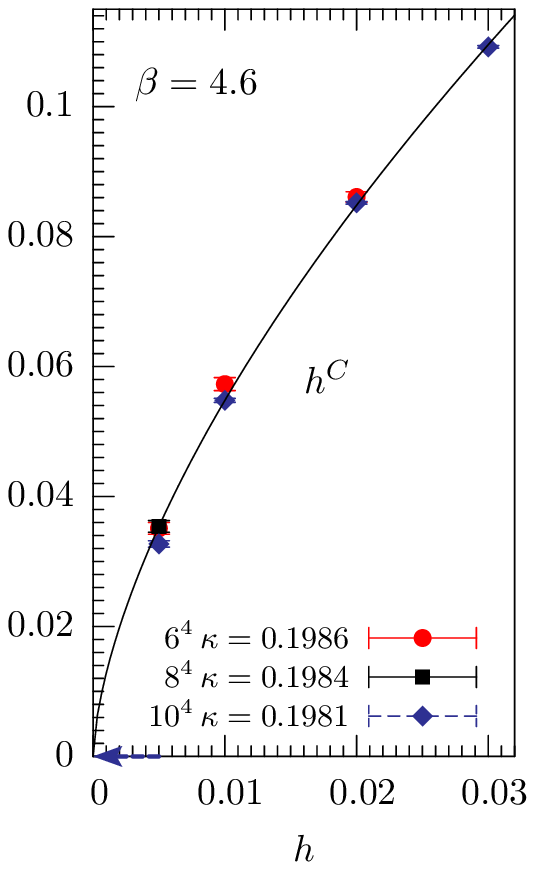}\hspace{-1cm}
    \includegraphics[height=8cm]{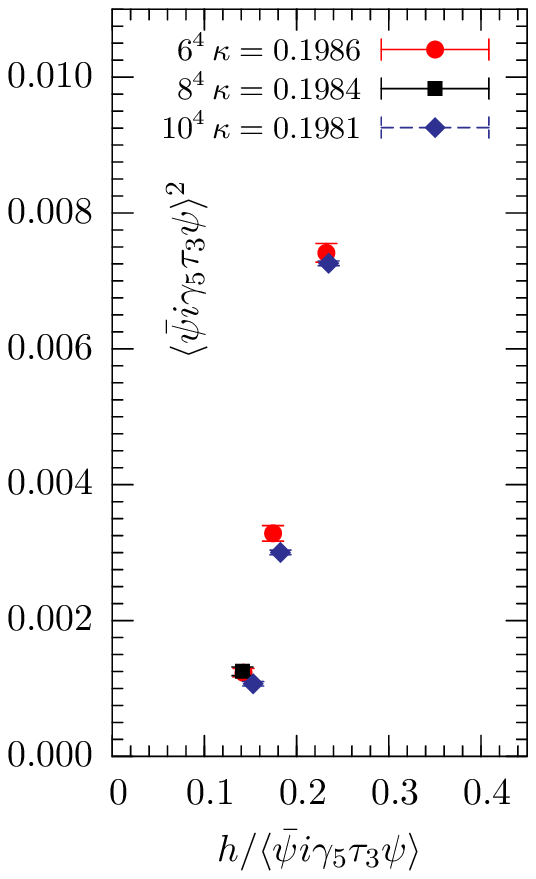}}
  \mbox{\includegraphics[height=8cm]{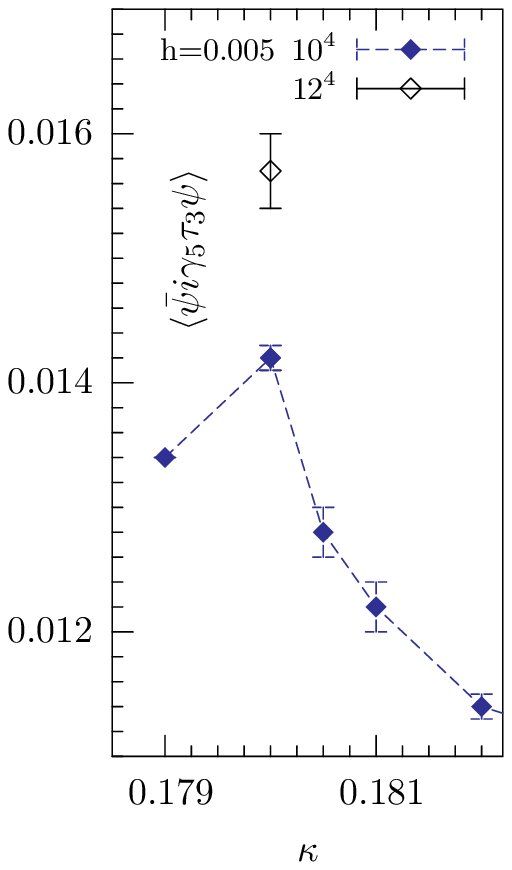}\hspace{-0.5cm}
    \includegraphics[height=8cm]{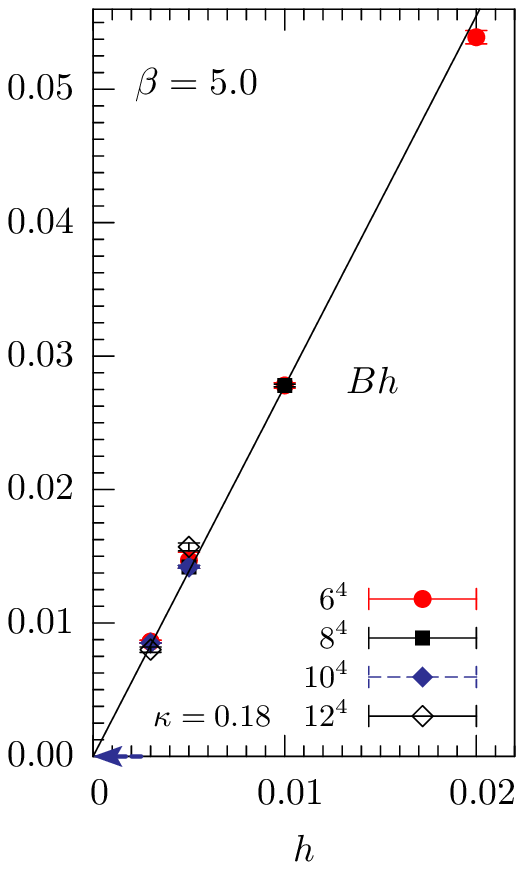}\hspace{-1.1cm}
    \includegraphics[height=8cm]{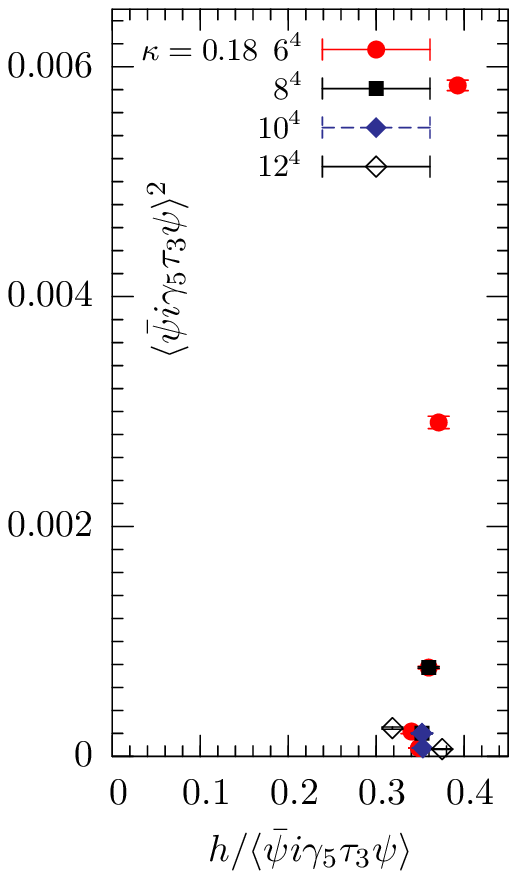}}
  \caption{In the left column data for $\op$ from lattices of various
  sizes are shown as a function of $\kappa$ at $h=0.005$ fixed
  (the lines are drawn to guide the eye).
  The extrapolation to $h=0$ in the infinite volume limit is
  shown in the center column.  In the right column the corresponding
  Fisher plots are shown. The upper row contains data at $\beta=4.6$
  from lattices of sizes $6^4$, $8^4$ and $10^4$.  The lower row shows
  measurements for $\beta=5.0$ from lattices of size from $6^4$ to
  $12^4$.}
\label{fig:extrapolation4650}
\vspace{0.3cm}
\end{figure*}

Using Eq.~(\ref{eq:ansatz}) with the mean field value 
$C=1/3$ results in unstable fitting 
functions, but taking $C$ as a free parameter instead, the ansatz
describes the data well. In fact, the parameter of interest $A$ is
robust against the introduction of linear and quadratic correction
terms (see Table \ref{tab:op_fit_404346}).  
Furthermore, the fit
parameters $B$ \mbox{and $C$} agree within errors for both values of $\beta$,
even when introducing corrections.  We conclude that the order
parameter $\op_{h=0}$ is non-zero at $(\beta,\kappa)=(4.0,0.22)$ and
$(4.3,0.21)$.

Measurements for $\op$ at $\beta=4.6$ are shown in the upper row of
Fig.~\ref{fig:extrapolation4650}.  Looking at the upper left plot of
the figure one sees that $\op$ still has a peak at finite $h$. 
The peak becomes narrower and its position is shifted from
$\kappa=0.1986$ to $\kappa=0.1981$ as the lattice size is increased
from $6^4$ to $10^4$. Taking the results from the $10^4$ lattice at
$\kappa=0.1981$, a fit using Eq.~(\ref{eq:ansatz}) can be performed.
However, due to low statistics the point at $h=0.005$ was discarded
and therefore some fit parameters had to be fixed. Using the fit
results from the two lower values of $\beta$, the parameter $A$, $B$ 
and $C$ were alternately fixed to reasonable values. The extrapolation 
is consistent with a vanishing order parameter (see 
Table~\ref{tab:op_fit_404346}). The same result is obtained by inspection 
of the Fisher plot in Fig.~\ref{fig:extrapolation4650}
where the data seem to lie on a line ending on the abscissa. 
This means that the order parameter $\op_{h=0}$ vanishes
at~$\beta=4.6$.

In addition, the parameters $B$ and $C$ agree within errors for all 
three values of $\beta$ as can be seen from Table~\ref{tab:op_fit_404346}.
Therefore, we also fitted the data globally using ansatz~(\ref{eq:ansatz})
where $B$ and $C$ are common to all data, while the parameters $A_{\beta}$
and $D_{\beta}$ are different for each $\beta$. In Table~\ref{tab:globfit}
the fit results are shown. In agreement with the results presented above,
the order parameter $\op_{h=0}$ is found to be finite at $\beta=4.0$ 
and $\beta=4.3$, while it vanishes at $\beta=4.6$.  Furthermore, 
their values are robust against the introduction of a correction 
term linear in $h$, while $B$ and $C$ are sensitive.

At $\beta=5.0$ a vanishing order parameter becomes manifest. 
As can be seen from the lower row of Fig.~\ref{fig:extrapolation4650} 
there is still a peak. However, the extrapolation of $\op$ to $h=0$ at 
$\kappa=0.18$ as well as the Fisher plot do not support a finite value 
of $\op_{h=0}$ at~$\beta=5.0$.

\begin{table*}
\renewcommand{\arraystretch}{1.0}       
\setlength{\tabcolsep}{4mm}             
\centering{\small 
  \begin{tabular}{|l||c|c|c|c|c||c|}
    
    \hline 
      fit  &     A     &   B      & C     &  D  &  E & \effchi \\
    \hline\hline 
           & \multicolumn{6}{|c|}{$\beta=4.0$\qquad$\kappa=0.22$}\\
    \hline\hline 
    \bf  1 & \bf0.068(4)& \bf1.07(9) & \bf0.67(3) &  \bf0  &  \bf0 &\bf0.80\\
    \hline
        2a &   0.067(3) & 1       & 0.66(2) & 0.1(1) & 0 & 0.83\\
        2b &   0.067(3) & 1.03(11) & $2/3$  & 0.1(3) & 0 & 0.81 \\
    \hline
        3a &   0.066(3) & 1       & 0.65(1) & 0 & 1(1)    & 0.98\\
        3b &   0.067(2) & 1.05(4) & $2/3$   & 0 & 0.1(17) & 0.84\\
    \hline\hline 
           & \multicolumn{6}{|c|}{$\beta=4.3$\qquad$\kappa=0.21$}\\
    \hline\hline 
    \bf  1 & \bf0.034(1) & \bf0.99(3) & \bf 0.65(1) & \bf0 & \bf 0 & \bf 0.08\\
    \hline
        2a &   0.034(1) & 1  & 0.65(1) & -0.02(4) & 0 & 0.08 \\
        2b &   0.035(1) & 1.11(4) & $2/3$ & -0.2(1) & 0 & 0.09 \\
    \hline
        3a &   0.035(1) & 1 & 0.65(1)  & 0 & -0.2(6) & 0.08 \\ 
        3b &   0.036(1) & 1.06(1) & $2/3$  & 0    & -1.2(9)  & 0.13 \\
    \hline\hline 
           & \multicolumn{6}{|c|}{$\beta=4.6$\qquad$\kappa=0.1981$}\\
    \hline\hline 
    \bf  1a  &  \bf 0      & \bf 1  & \bf 0.63(1)  & \bf 0  & \bf 0 &\bf 3.21\\ 
         1b  &  0.001(2)   & 1      & 0.63(6)      & 0      & 0     & 4.82\\
         1c  &  0          & 0.97(3)& 0.62(9)      & 0      & 0     & 3.62\\
    \hline
          2a  &  0          & 1      & 0.63(1)  & -0.05(5)  & 0     & 3.36\\
          2b  &  0.0005(8)  & 1      & 0.63     & -0.03(3)  & 0     & 3.87\\
    \hline
          3a &  0          & 1      & 0.63(1)  & 0         & -1(1)   & 2.41\\
          3b &  0.0003(4)  & 1      & 0.63     & 0         & -0.9(7) & 2.79\\
    \hline 
  \end{tabular}}
  \caption{The parameters of the ansatz $\sigma(h)=A+Bh^{C}+Dh+Eh^2$ fitted
           to the data for $\op$ at $\beta=4.0$, $4.3$ and
           $4.6$ with no, linear or quadratic corrections 
           (labeled as 1, 2 or 3). At each $h$ the result from the 
           largest lattice was used in the fit (for details
           see Table \ref{tab:stat}). The data at $h=0.003$ 
           were discarded because these are from a $6^4$ lattice. 
           Also the result at $\beta=4.6$ and $h=0.005$ was not taken
           into account due to low statistics.
           Fixed parameters are presented by their value without 
           giving an error. In each case the first fit
           (bold numbers) was used in Figs.~\ref{fig:extrapolation4043} 
           and \ref{fig:extrapolation4650}, respectively.}
  \label{tab:op_fit_404346}
\end{table*}
\nopagebreak

\begin{table*}
\renewcommand{\arraystretch}{1.0}       
\setlength{\tabcolsep}{2mm}             
\centering{\small
\begin{tabular}{|c|c|r|r|r|r|r|r|r|r|r|r|}
 \hline 
         &  &            \multicolumn{10}{|c|}{$h$} \\
 \cline{3-12} 
    \raisebox{1.5ex}[-1.5ex]{$\beta$} & \raisebox{1.5ex}[-1.5ex]{$\kappa$} &
                         \multicolumn{2}{|c|}{$0.005$}  & 
                         \multicolumn{2}{|c|}{$0.01$}   &     
                         \multicolumn{2}{|c|}{$0.02$}   & 
                         \multicolumn{2}{|c|}{$0.03$}   & 
                         \multicolumn{2}{|c|}{$0.04$} \\
 \hline 
 \hline 
  4.0 & 0.2200 & $8^4$ & 250 & $10^4$& 146 & $6^4$ & 1000 & -- & -- & $6^4$ & 1000\\
  4.3 & 0.2100 & $8^4$ & 300 & $8^4$ & 500 & $8^4$ & 250  & $6^4$ & 500 & --  & --\\
  4.6 & 0.1981 &  --   &  -- & $10^4$& 170 & $10^4$& 250  & $10^4$& 200 & --  & --\\
 \hline 
\end{tabular}}
\caption{Statistics used for the final analysis (extrapolation) at selected 
  $\kappa$ values for $\beta=4.0$, $4.3$ and $4.6$. For the respective 
  values of $h$ given in the first row in the second column we report 
  the number of trajectories produced for each lattice size. A similar 
  statistic was used for scanning at neighboring $\kappa$ values.}
\label{tab:stat}
\end{table*}


\begin{table*}
\renewcommand{\arraystretch}{1.0}       
\setlength{\tabcolsep}{4mm}             
\centering{\small 
  \begin{tabular}{|l||c|c|c|c|c||c|}
    \hline 
      fit & $\beta$  &   A$_{\beta}$   &   B & C   & D$_{\beta}$ &\effchi \\
    \hline\hline 
          &  4.0     &    0.063(2)    &         &         & 0  & \\
        1 &  4.3     &    0.032(2)    & 1.0(1) & 0.64(2) & 0 & 3.4 \\
          &  4.6     &    0.004(2)    &         &         & 0  & \\ 
    \hline 
          &  4.0     &    0.065(2)    &         &         & -0.8(3) & \\ 
        2 &  4.3     &    0.034(2)    & 1.5(2)  & 0.71(2) & -0.8(3) & 2.4\\
          &  4.6     &    0.004(2)    &         &         & -0.7(3) & \\
    \hline 
  \end{tabular}}
  \caption{The parameters of the ansatz $\sigma(h)=A_{\beta}+Bh^{C}+D_{\beta}h$
           \, fitted to the data in Table~\ref{tab:stat}
           for $\op$ at $\beta=4.0$, $4.3$ and $4.6$ with 
           no (fit 1) and linear (fit 2) corrections. The parameter $B$ and 
           $C$ are common to all data, while for each $\beta$ there is a
           separate value for $A_{\beta}$ and $D_{\beta}$, respectively. 
           Fixed parameters are presented by their value without giving 
           an error.}
  \label{tab:globfit}
\end{table*}


\section{Discussion} \label{sec:discussion}

\begin{figure*}
  \centering
  \includegraphics[height=9cm]{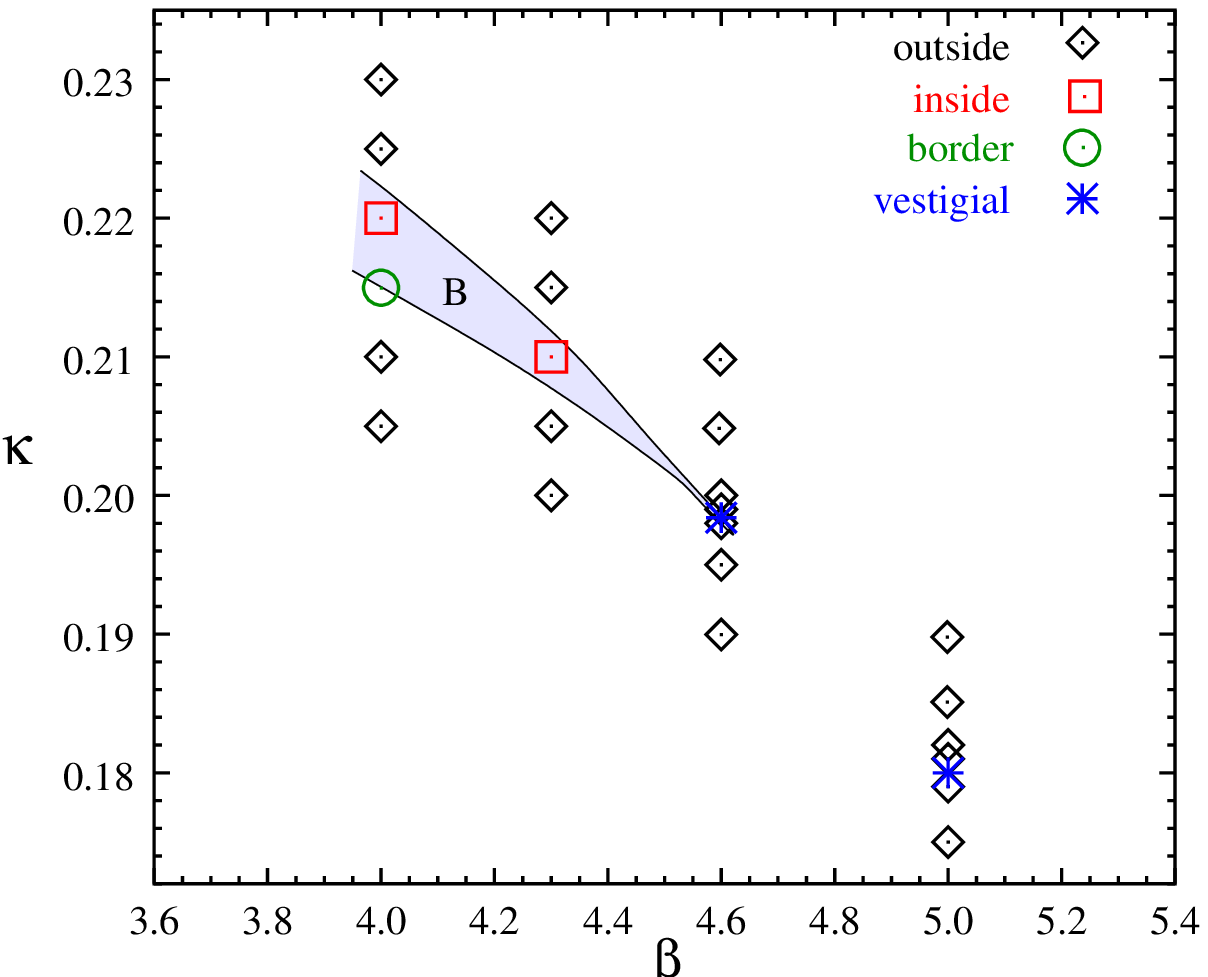}
  \caption{The part of the phase diagram studied in this work. Squares
    denote ($\beta,\kappa$) pairs where a finite value of $\op_{h=0}$
    is found.  Diamonds refer to points where $\op_{h=0}=0$. Stars
    denote points where a finite value of $\op_{h=0}$ is uncertain.
    The lines indicate the position of the critical lines
    $\kappa^{(l)}_c(\beta)$ and $\kappa^{(u)}_c(\beta)$. The shaded
    region labeled $B$ refers to the parity-flavor breaking
    phase. The point ($\beta,\kappa$)=($4.0,0.215$) marked by a circle 
    seems to lie very close to the border of the broken phase.}
 \label{fig:phasediagram_results}
\end{figure*}

In this study we have investigated how far a parity-flavor breaking
phase in lattice QCD with two flavors of dynamical Wilson fermions at
zero temperature extends in $\beta$.  An explicitly symmetry breaking
term, the twisted mass term $h\bar{\psi}i\gamma_{5}\tau^{3}\psi$, was
added to the Wilson fermion matrix.  The phase diagram was explored in
the rectangle $4.0\leq\beta\leq5.0$ and $0.15\leq\kappa\leq0.28$.

We have presented hybrid Monte Carlo results for the order parameter
$\op$.  The existence of a parity-flavor breaking phase could be
confirmed at \mbox{$(\beta,\kappa)=(4.0,0.22)$} and $(4.3,0.21)$, where
$\op$, measured at finite $h$, extrapolates to a finite value at $h=0$
in the infinite volume limit.  No parity-flavor breaking was found at
$\beta=4.6$ and $\beta=5.0$. This suggests a phase structure as shown
in Fig.~\ref{fig:phasediagram_results}.  Two squares in
Fig.~\ref{fig:phasediagram_results} mark points where we were able to
confirm the Aoki phase.  Two stars mark points where $\op$ has a peak
at finite $h$, but where our extrapolation to $h=0$ is consistent with
a vanishing order parameter. Consequently these stars are labeled
\emph{vestigial}.

According to these results, the Aoki phase for $T=0$ seems to end close to 
$\beta=4.6$ and  $\kappa=0.1981$. 
Rough estimates for the upper $\kappa^{(u)}_c$ and 
lower $\kappa^{(l)}_c$ bound of the Aoki phase are
\begin{eqnarray*}
        &\beta=4.0:&\; 0.215\simeq\kappa^{(l)}_c<0.220
                    \quad 0.220<\kappa^{(u)}_c<0.225 \nonumber\\
        &\beta=4.3:&\; 0.205<\kappa^{(l)}_c<0.210
                    \quad 0.210<\kappa^{(u)}_c<0.215\,.
\end{eqnarray*}
The pair $(\beta,\kappa) = (4.0,0.215)$ seems to be quite close to the
lower boundary.  We conclude this from the behavior of $\op$ in conjunction 
with the behavior of the pion norm \cite{BKR} (which also has been measured
during our simulations).  At this $(\beta,\kappa)$ pair $\op$
extrapolates to zero at $h=0$, whereas the pion norm seems to diverge
as $h\rightarrow0$.  Such behavior is expected close to critical lines
$\kappa_c(\beta)$.

Referring to the discussion of the anticipated phase diagram in
Sec.~\ref{sec:Aokiphase}, the results presented here do not
indicate a parity-flavor breaking phase at $\beta \ge 4.6$ which was
originally claimed to exist at all $\beta$ (see
Fig.~\ref{fig:proposed_phase_diagram}). This is suggested not only by
the extrapolation of $\op$ to $h=0$ in Sec.~\ref{sec:extrapolation},
which yields $\op_{h=0}=0$ at $\beta=4.6$ and $\beta=5.0$, but also by
the observation that the peak of $\op$ decreases in height and width
as $\beta$ increases.  The parity-flavor breaking phase seems to be 
pinched off near $\beta=4.6$ as illustrated in
Fig.~\ref{fig:phasediagram_results}.  From the numerical point of view
we agree with Bitar \cite{BitPRD}, who has found no evidence of such a
broken phase for $\beta\ge5.0$. 

On the other hand, although $\op$ decreases, a non-vanishing value at $h=0$
in the infinite volume limit is not excluded.  A decreasing width
could have been expected from the phase structure in
Fig.~\ref{fig:proposed_phase_diagram}.  The fact that the peak becomes
narrower implies that a high resolution scan in $\kappa$ is required
at larger values of $\beta$.  In addition, lattices much
larger than~$12^4$ would be needed
(which is beyond our presently available computing resources). 
With this in mind it is 
comprehensible that the results presented in Ref.~\cite{BitPRD} 
could not indicate a broken phase for $\beta\ge5.0$ just because 
of the small lattice sizes used ($6^4$, $8^4$ and $10^4$).
While we find no numerical evidence for the existence of the Aoki phase 
for $\beta\ge4.6$ one cannot exclude that the phase might be found
with methods to be invented similar to reweighting.

A further interesting observation we made is that the 
data behave differently when approaching the 
parity-flavor breaking phase at fixed $\beta$ from
$\kappa > \kappa^{(u)}_c$  compared with the approach
from $\kappa < \kappa^{(l)}_c$.  First, the peaks of $\op$ as a
function of $\kappa$ are asymmetric.  Second, an autocorrelation
analysis of $\op$ shows that measurements to the right of the peak
(above of the Aoki phase) are significantly 
stronger correlated than at all smaller $\kappa$ values.

In light of the possible scenarios discussed by Sharpe and 
Singleton~\cite{ShaSin}
it might be worthwhile to invest more computing power in a study of
both the width of the \emph{Aoki finger} and the detailed behavior of the 
neutral pion mass inside and outside the Aoki phase with respect to the 
lattice spacing dependence. For the case of $N_f=2$ dynamical (unimproved) 
Wilson fermions with standard Wilson gauge action, however, the impossibility 
of matching the $\pi$ and $\rho$ masses in the interval $3.5 < \beta < 5.3$
is known~\cite{BitEdwHelKen}, which means that scaling is strongly violated. 
Thus the result of the present paper, confining the Aoki phase to 
$\beta < 4.6$, unfortunately does not allow this potentially interesting 
comparison with chiral perturbation theory. 

In view of the fact that the region above the Aoki phase 
($\kappa > \kappa^{(u)}_c$) is the region
of interest for the insertion of the Wilson-Dirac operator into the
overlap form \cite{Neu} of the massless fermion operator an 
even more extensive investigation of this area might be 
worthwhile to do.

\newpage
\section*{ACKNOWLEDGMENTS}

All simulations were done on the Cray T3E at  
Konrad-Zuse-Zentrum f\"ur Infor\-mations\-technik Berlin. 
A. S. would like to thank the DFG-funded graduate 
school GK~271 for financial support. \mbox{E.-M.~I.} and 
\mbox{M.~M.-P.} acknowledge 
support from the DFG research group ``Hadron Lattice Phenomenology''
FOR 465.



\end{document}